\documentclass[lettersize,journal]{IEEEtran}
\usepackage{amsmath,amsfonts}
\usepackage{algorithmic}
\usepackage{algorithm}
\usepackage{array}
\usepackage[caption=false,font=normalsize,labelfont=sf,textfont=sf]{subfig}
\usepackage{textcomp}
\usepackage{stfloats}
\usepackage[hyphens]{url}
\usepackage{verbatim}
\usepackage{graphicx}
\usepackage{cite}
\usepackage[colorlinks=true,linkcolor=black,anchorcolor=black,citecolor=black,filecolor=black,menucolor=black,runcolor=black,urlcolor=black,breaklinks=true]{hyperref}
\usepackage{booktabs}
\usepackage{enumitem}

\usepackage{background}
\backgroundsetup{
  position=current page.north west,
  angle=0,
  nodeanchor=west,
  vshift=-5mm,
  hshift=2.5cm,
  color=black,
  opacity=1,
  scale=0.7
}
\SetBgContents{\parbox{25cm}{
This article has been accepted for publication in IEEE Transactions on Visualization and Computer Graphics. This is the author's version which has not been fully edited and content may change prior to final publication. Citation information: DOI 10.1109/TVCG.2024.3410097
}}

\begin{document}

\newcommand{\demolink}[1]{\href{https://responsive-vis.github.io/breakpoints/demos/#1}{responsive-vis.github.io/breakpoints/demos/#1}}
\newcommand{\sitelink}{\href{https://responsive-vis.github.io/breakpoints/}{responsive-vis.github.io/breakpoints}}

\newcommand{\viewsfirst}{\textbf{\textsc{View-led}}}
\newcommand{\conditionsfirst}{\textbf{\textsc{Constraint-led}}}
\newcommand{\mixed}{\textbf{\textsc{Mixed}}}

\title{%
Constraint-Based Breakpoints for Responsive Visualization Design and Development
}

\author{Sarah Schöttler, Jason Dykes, Jo Wood, Uta Hinrichs, and Benjamin Bach
\thanks{Sarah Schöttler (\texttt{\href{mailto:sarah.schoettler@ed.ac.uk}{sarah.schoettler@ed.ac.uk}}) and Uta Hinrichs (\texttt{\href{mailto:uhinrich@ed.ac.uk}{uhinrich@ed.ac.uk}}) are with the University of Edinburgh, UK.}
\thanks{Benjamin Bach (\texttt{\href{mailto:bbach@ed.ac.uk}{bbach@ed.ac.uk}}) is with Inria, France and the University of Edinburgh, UK.}%
\thanks{Jason Dykes (\texttt{\href{mailto:j.dykes@city.ac.uk}{j.dykes@city.ac.uk}}) and Jo Wood (\texttt{\href{mailto:j.d.wood@city.ac.uk}{j.d.wood@city.ac.uk}}) are with City, University of London, UK.}}
% \author{IEEE Publication Technology,~\IEEEmembership{Staff,~IEEE,}
%         % <-this % stops a space
% \thanks{This paper was produced by the IEEE Publication Technology Group. They are in Piscataway, NJ.}% <-this % stops a space
% \thanks{Manuscript received April 19, 2021; revised August 16, 2021.}}

% The paper headers
\markboth{Journal of \LaTeX\ Class Files,~Vol.~14, No.~8, August~2021}%
{Shell \MakeLowercase{\textit{et al.}}: A Sample Article Using IEEEtran.cls for IEEE Journals}

% \IEEEpubid{0000--0000/00\$00.00~\copyright~2021 IEEE}
% Remember, if you use this you must call \IEEEpubidadjcol in the second
% column for its text to clear the IEEEpubid mark.

\maketitle

\begin{abstract}
    This paper introduces constraint-based breakpoints, a technique for designing responsive visualizations for a wide variety of screen sizes and datasets.
Breakpoints in responsive visualization define when different visualization designs are shown. 
Conventionally, breakpoints are static, pre-defined widths, and as such do not account for changes to the visualized dataset or visualization parameters. 
To guarantee readability and efficient use of space across datasets, these static breakpoints would require manual updates. 
Constraint-based breakpoints solve this by evaluating visualization-specific constraints on the size of visual elements, overlapping elements, and the aspect ratio of the visualization and available space. 
Once configured, a responsive visualization with constraint-based breakpoints can adapt to different screen sizes for any dataset. 
We describe a framework that guides designers in creating a stack of visualization designs for different display sizes and defining constraints for each of these designs.
We demonstrate constraint-based breakpoints for different data types and their visualizations: geographic data (choropleth map, proportional circle map, Dorling cartogram, hexagonal grid map, bar chart, waffle chart), network data (node-link diagram, adjacency matrix, arc diagram), and multivariate data (scatterplot, heatmap). 
Interactive demos and supplemental material are available at \sitelink.

\end{abstract}

\begin{IEEEkeywords}
Information visualization, responsive visualization
\end{IEEEkeywords}

\section{Introduction}
\label{sec:intro}

Visualizations are increasingly being accessed from a variety of devices such as smart watches, smartphones, tablets, monitors, and wall-sized displays.
These devices come with many different screen sizes, resolutions, aspect ratios, input and interaction modalities, hardware and network capabilities, and usage contexts \cite{chittaroVisualizingInformationMobile2006,horakResponsiveVisualizationDesign2021}. 
Likewise, the display space available for a visualization can suddenly change, e.g., when users rotate their device, resize a window, or modify the layout in a multi-view application.
Responsive visualization design aims to create visualizations that adapt to these different display sizes and device contexts, ensuring that visualizations do not become too small to be read, suffer from overlapping elements, nor have unsuitable aspect ratios.
This process of designing responsive visualizations requires navigating trade-offs around visual information density and optimizing legibility while maintaining the \textit{message}~\cite{kimDesignPatternsTrade2021} of a visualization.

Currently, visualization designers often create different versions of a visualization manually, optimized for different sizes and possibly interaction modalities~\cite{kimDesignPatternsTrade2021, hoffswellTechniquesFlexibleResponsive2020}. 
However, which version a viewer will see for a given display size is largely determined by a single value: the \textit{viewport width}, i.e., the width of the browser window or screen. 
In practice, a width of 768px is a commonly used threshold---called a \textit{breakpoint}---below which the mobile-focused version of a visualization is shown~\cite{w3schoolsResponsiveWebDesign}.
While such \textit{device breakpoints} are an established approach in responsive web design, they limit reusability and reliability of responsive visualizations across datasets, devices, and usage scenarios. Device breakpoints need to be reviewed and possibly reconfigured each time the dataset is changed, updated, or replaced. In applications with quick and unpredictable changes to the data (e.g., real-time dashboards or trackers, live news visualizations, or user-filterable visualizations), manually reviewing and adapting visualization designs and breakpoints is not feasible.

We introduce \textbf{constraint-based breakpoints} which automatically guarantee high legibility and usability of a visualization for \textit{every} display size, dataset, and visualization design.
With constraint-based breakpoints, designers assign each version of the visualization one or more constraints that ensure its legibility, such as a minimum size for visual elements, a maximum level of overlap between visual elements, or a limit to its aspect ratio.
Once a responsive visualization has been configured in this way, constraint-based breakpoints dynamically determine the most suitable version of the visualization to display by evaluating each version's constraints for the current display size and dataset.
For designers who want to adopt this approach, we suggest two complementary workflows and introduce the view landscape, a visualization illustrating the effect of setting different constraints.

To facilitate creating responsive visualizations with constraint-based breakpoints, 
we develop a proof-of-concept JavaScript library. We use this library to create four example visualizations, covering a range of data types (geographic, network, multivariate) and visualization designs (maps, cartograms, node-link diagrams, scatterplots, and more). Our examples demonstrate the process and results of designing with constraint-based breakpoints, including how to set \textit{view constraints} for a \textit{view stack} of different visualization designs. Interactive demos of all our examples, their code, and our library are available at \sitelink.

\section{Related Work}
\label{sec:rw}

Responsive visualization requires solutions to two separate yet intertwined issues: \textit{i)} creating visualization designs for different viewport sizes (Sec.~\ref{sec:rw:designs}), and \textit{ii)} defining rules about which design to show to each viewer (Sec.~\ref{sec:rw:deciding}). 

\subsection{Creating Visualization Designs}
\label{sec:rw:designs}

Responsive visualization designers commonly start with one visualization design, e.g., for a desktop screen, and then adapt this design to other screen sizes~\cite{hoffswellTechniquesFlexibleResponsive2020}. To support this process, designers can rely on a range of dedicated tools and guidelines.

Prior work has classified \textbf{strategies and design patterns} for responsive visualization used by practitioners in news contexts. 
Kim~et~al.~\cite{kimDesignPatternsTrade2021} introduce a taxonomy of design patterns and strategies for responsive visualizations based on a sample of 378 pairs of small and large screen visualizations collected from online news media and similar sources. 
Hoffswell~et~al.~\cite{hoffswellTechniquesFlexibleResponsive2020} describe a set of strategies, such as `remove title' or `reposition labels', obtained from reviewing 231 responsive visualizations. 
For coordinated multiple views (CMV) visualizations~\cite{wangbaldonadoGuidelinesUsingMultiple2000,robertsStateArtCoordinated2007}, a design space and preliminary design guidelines have been proposed based on a user study comparing three different views of a CMV visualization~\cite{badamEffectsScreenresponsiveVisualization2021}.

\textbf{Authoring tools and libraries} support the design process more directly.
Hoffswell~et~al.~\cite{hoffswellTechniquesFlexibleResponsive2020} describe an authoring system that propagates design changes across multiple views, allowing designers to simultaneously work on multiple designs for a range of sizes. 
Kim~et~al.'s \textit{declarative grammar} `Cicero' allows designers to create multiple views at different sizes by creating a source view and specifying transformations to create one or more target views, which they implemented as a library on top of Vega-Lite \cite{kimCiceroDeclarativeGrammar2022}.
`Dupo' is a \textit{mixed-initiative} tool that automatically generates resized versions of an initial design, then lets the user choose from and modify these designs with more manual authoring tools~\cite{kimDupoMixedInitiativeAuthoring2024}. 

\textbf{Automated approaches} aim to redesign visualizations for different sizes without requiring a designer's input (or significantly reducing their input). 
Several automations aim to target a whole range of visualization types by making generic layout changes, such as rearranging and rescaling elements. 
\textit{Grid-based resizing} can be applied to word clouds, node-link diagrams, and other visualization techniques without meaningful x- and y-positions~\cite{wuViSizerVisualizationResizing2013}. 
For line charts, bar charts, scatterplots, and other cartesian charts, one technique uses \textit{reinforcement learning} to achieve automatic resizing~\cite{wuMobileVisFixerTailoringWeb2021}; another technique uses a \textit{recommender engine} to generate several candidate visualizations and rank their preservation of `task-oriented insights' using a loss function~\cite{kimAutomatedApproachReasoning2022}.
The loss function in this engine targets similar measures as some of our \textit{view constraints}, though our constraints are used at runtime to determine the most suitable view out of the available options, rather than to rank different options during the design process.

The problem of making best use of limited screen space is further well known in areas of visualization concerned with complex data such as \textbf{geographic, network, and multivariate data}. 
In geographic visualization, map generalization is the practice of adapting maps for different scales by `representing various geographies at different levels of detail'~\cite[p.~1]{stanislawskiGeneralisationOperators2014}.
There are a variety of techniques---called \textit{generalization operators}~\cite{rothTypologyOperatorsMaintaining2011}---to achieve this, such as line simplification algorithms, automated displacement of elements or labels, or strategies such as typification or symbolization. 
Although map generalization research has largely focused on topographic maps as opposed to thematic maps in information visualization, generalization operators are used for thematic maps as well~\cite{raposoChangeThemeRole2020}. 
Setlur and Chung~\cite{setlurSemanticResizingCharts2021} further argue that generalization techniques could be applied to non-geographic charts and demonstrate a simplification technique for responsive line charts based on map generalization techniques. 
For network data, different layout algorithms for node-link diagrams can cater to different aspect ratios and sizes~\cite{gibsonSurveyTwodimensionalGraph2013}. 
Likewise, matrices are good for dense networks in cases where certain topological patterns matter, but do not scale well with the number of nodes for other patterns. 
For extreme aspect ratios, triangle matrices (e.g.,~\cite{baudrySerpentineFlexible2D2020} Fig.~1) or arc diagrams may fit for specific datasets.

Constraint-based breakpoints complement these existing tools and strategies by providing designers with fine-grained control over what users see across a range of devices. 
For example, a designer could use any of the above authoring or automation techniques to create the different views of their visualization, then set constraint-based breakpoints between these versions to determine which version each user sees.

\subsection{Deciding Which Visualization Design to Show}
\label{sec:rw:deciding}

In \textbf{responsive web design}, the current best practice is to manage changes across different devices with \textit{device breakpoints}. 
Device breakpoints typically reference different viewport widths and are implemented using CSS Media Queries~\cite{w3schoolsResponsiveWebDesign}. 
Many web frameworks and libraries come with pre-defined breakpoints and provide developers with categories ranging from, e.g., extra-small to extra-large, representing specific viewport width ranges (e.g.~\cite{tailwindcontributorsResponsiveDesignTailwind2023,bootstrapcontributorsBreakpoints2023}). 
In the context of visualization, Horak~et~al.~\cite{horakResponsiveVisualizationDesign2021} call for a \textit{responsive design mindset} that considers responsive design as an integral part of any visualization design process.
However, visualization design poses additional challenges compared to generic web design. As established by Walny~et~al.~\cite{walnyDataChangesEverything2020}, \textbf{visualization design processes} must consider factors not present in generic design processes, and tools are often lacking in this regard. 
Five types of factors that responsive visualizations should flexibly account for are suggested by Horak~et~al.~\cite{horakResponsiveVisualizationDesign2021}:
\textit{device factors} (e.g. display size, interaction modalities), \textit{usage factors} (e.g. orientation of device), \textit{environmental factors} (e.g. lighting situation), \textit{data factors} (e.g. structure, size), and \textit{human factors} (e.g. visualization literacy, tasks).

Yet, \textbf{practical approaches and tools} that support managing different versions of a visualization in a way that considers factors beyond the viewport width are currently sparse. 
Early work on responsive visualization design includes R3S.js~\cite{leclaireR3SJsResponsive2015}, a responsiveness-focused extension of D3.js that includes features such as pre-defined media queries that consider the orientation of a device in addition to the screen size. 
\textit{Gosling}~\cite{lyiGoslingGrammarbasedToolkit2022}, a toolkit for genomics data visualization, lets designers modify or hide visualizations based on the height, width, and zoom level~\cite{lyiMultiViewDesignPatterns2022}. 
For specific visualization types (line charts, bar charts, parallel coordinate plots, and scatter plots), proof-of-concept responsive visualizations with custom resizing rules and breakpoints have been demonstrated~\cite{andrewsResponsiveVisualisation2018}. 
Due to how labor-intensive this manual process is, custom rules for resizing visualizations are currently rare outside of research prototypes as described here.
Our work essentially generalizes and extends this manual approach by providing a structured way of setting custom, visualization-specific constraints. 
As such, constraint-based breakpoints are a first foray into making some of Horak~et~al.'s ideas on flexibly adapting to various factors more achievable in practice. 

\section{Motivating Example}
\label{sec:motivating-example}

\begin{figure}
    \centering
    \includegraphics[width=\columnwidth]{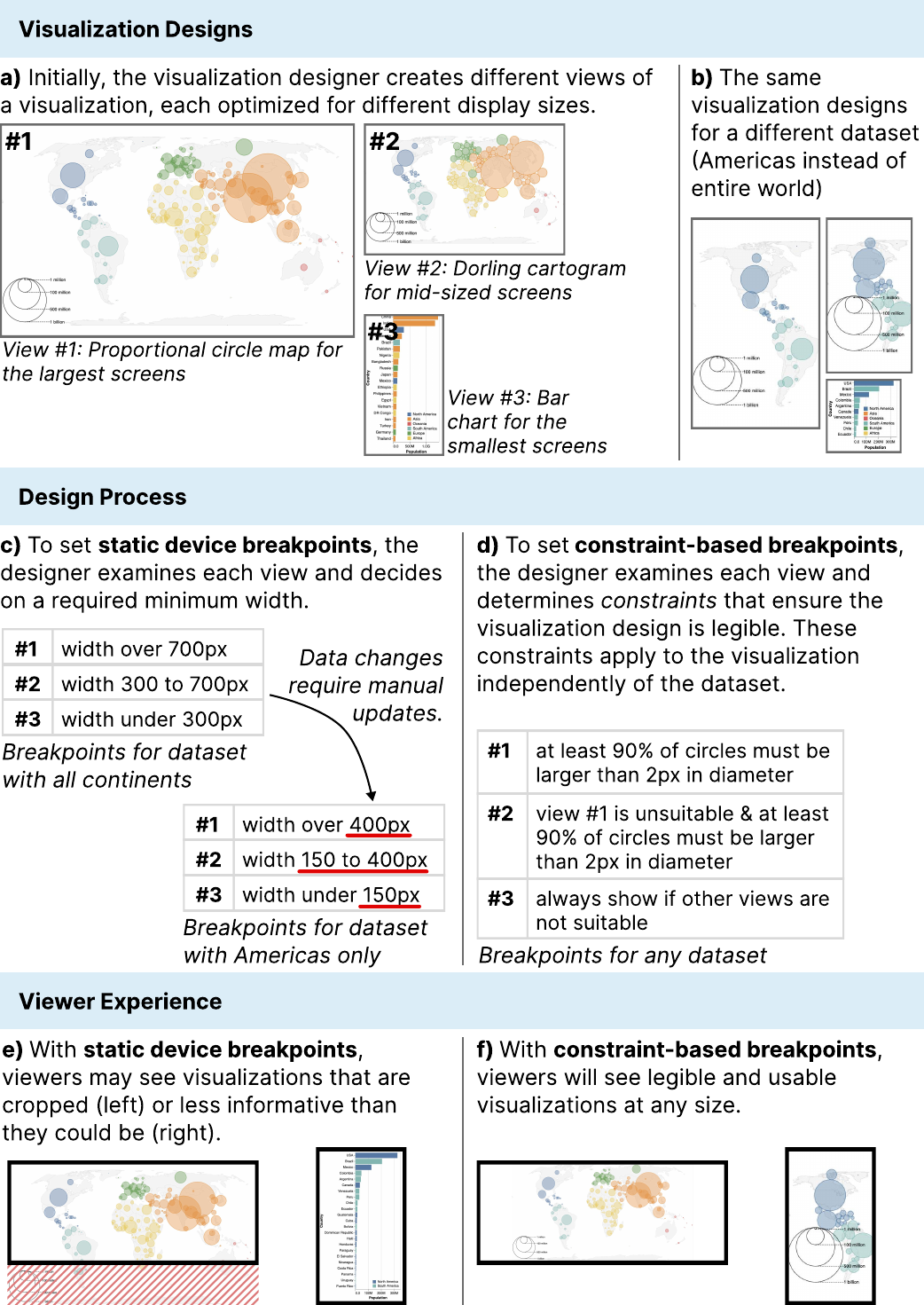}
    \caption{Motivating example demonstrating problems with conventional device breakpoints from designer and viewer perspectives, and how they are addressed by constraint-based breakpoints.}
    \label{fig:example}
\end{figure}

This section gives an example of how constraint-based breakpoints work and illustrates which problems they solve compared to conventional device breakpoints. 
Figure~\ref{fig:example}a shows three possible visualization designs for displaying global population by country, optimized for large, medium, and small display sizes. 
A proportional circle map (\#1) provides detailed geographic context but when scaled down, circles will become too small (when the map and circles are scaled together) or overlap (when the map is scaled but circle scale kept constant). 
To address these problems for smaller sizes, a Dorling cartogram (\#2) can be a good alternative: it prevents circles from overlapping by displacing them, allowing for larger circles by sacrificing geographic detail.
For the smallest viewport sizes, a world map may not reasonably fit at all, so a space-efficient bar chart (\#3) is chosen. 
As an example for filtering or updating the visualized dataset, we also show the same visualization designs for a data subset containing just the Americas in Fig.~\ref{fig:example}b.

Using conventional \textbf{device breakpoints}, 
the visualization \textbf{designer} has to examine each view and determine at which width it becomes too small. Then, they  set breakpoints accordingly (Fig.~\ref{fig:example}c). 
This process is relatively easy to manage and well-supported by web development frameworks, but needs to be repeated whenever the underlying data---and therefore the visualization---changes. For example, when showing only the Americas (Fig.~\ref{fig:example}b) rather than all continents, the map has a narrower aspect ratio, and there is a smaller number of circles with a different size distribution, which will require breakpoints at different widths. 
From a \textbf{viewer} perspective, device breakpoints work well on most devices. However, since they usually only account for the viewport width, viewers who use unexpected screen or window sizes may experience cropped visualizations or unused white space that could have been used to display the data more clearly (Fig.~\ref{fig:example}e). 
Viewers might also experience these issues when the visualization's aspect ratio or data distribution change, e.g., through real-time data updates or interactive filtering.

With \textbf{constraint-based breakpoints}, the \textbf{designer} uses our library (Sec~\ref{sec:breakpoints:implementation}) to set constraints that ensure legibility and usability for each of the three initial visualization designs (Fig.~\ref{fig:example}d): for the proportional circle map (\#1) and the Dorling cartogram (\#2), \textit{at least 90\% of circles in the visualization must have a diameter of at least 2px.}
The designer also decides that \#1 is the preferred view; once its constraint is not satisfied anymore, \#2 should be displayed, and only if the constraint for \#2 fails as well, \#3 should be used as a last option.
Now, if the data changes, resulting in a different aspect ratio or distribution of circles on the map, our library automatically re-evaluates the constraints and updates the visualization.
For example, evaluating the constraint for the proportional circle map (\#1) for the full global dataset results in a minimum size of approximately 600$\times$320 pixels.
Applying the \textit{same constraint} to just the Americas data yields a much smaller lower limit on the display size of approximately 300$\times$380 pixels. 
\textbf{Viewers} will now always see a legible and usable visualization, even with unusual display sizes and frequent or unpredictable changes to the data (Fig.~\ref{fig:example}f).
We discuss this example in more detail in Sec.~\ref{sec:examples} but first introduce our toolkit for constraint-based breakpoints.

\section{Constraint-Based Breakpoints}
\label{sec:breakpoints}

The following subsections introduce our toolkit for designing with constraint-based breakpoints.
Sec.~\ref{sec:breakpoints:vl} introduces the \textbf{view landscape} as a visual tool for understanding the effects of different breakpoints and constraints.
As a structured framework for creating visualizations with constraint-based breakpoints, we then introduce the \textbf{view stack and view constraints} (Sec.~\ref{sec:breakpoints:stack}).
We also suggest two complementary \textbf{workflows} for designers (Sec.~\ref{sec:breakpoints:workflows}) and introduce our \textbf{prototype library} (Sec.~\ref{sec:breakpoints:implementation}) for constraint-based breakpoints.

\subsection{View Landscapes}
\label{sec:breakpoints:vl}

\begin{figure}
    \centering
    \includegraphics[width=1\columnwidth]{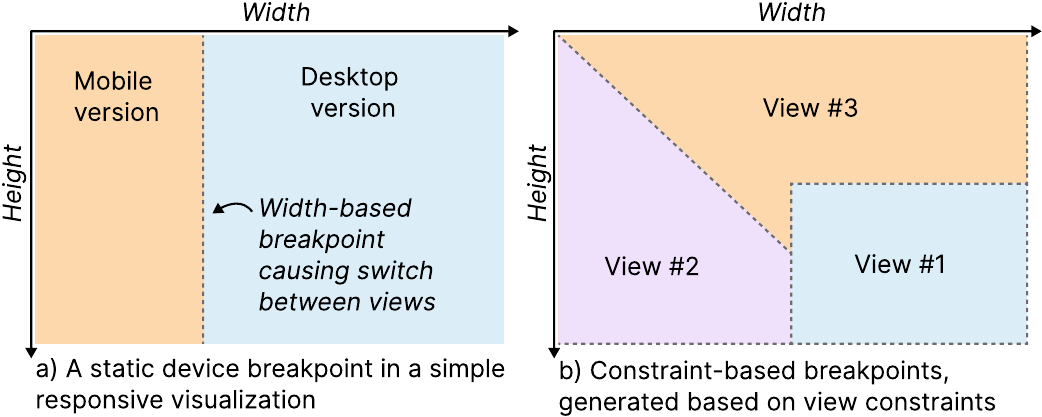}
    \caption{\textbf{View landscapes} for (a) a simple responsive visualization with two views and a fixed-width device breakpoint and (b) a responsive visualization with constraint-based breakpoints, illustrating which views (different colors) are shown for each width-height combination.}
    \label{fig:view-landscapes}
\end{figure}

We use diagrams that we call \textit{view landscapes} to illustrate how constraint-based breakpoints work, and how they differ from generic device breakpoints. 
A view landscape (Fig.~\ref{fig:view-landscapes}) maps out the space of all possible width-height combinations with width on the x-axis and height on the y-axis. Each pixel, representing each possible combination of width and height, is shaded by what view of the visualization is displayed at that size. The view landscape is a conceptual tool, intended to help create, describe, and evaluate responsive visualizations. 
While constraint-based breakpoints make it easier to reliably respond to a wider range of screen sizes and dataset changes, they can be more challenging to test and debug than conventional device breakpoints. In this context, the view landscape acts as a tool to help designers understand the effect of different constraints, and which views will be visible at what sizes.
View landscapes are inspired by diagrams used in cartography to provide an overview of different symbolizations and map projections across a range of map scales (e.g.,~\cite{brewerFramingGuidelinesMultiScale2007,jennyAdaptiveCompositeMap2012}).

Figure~\ref{fig:view-landscapes} shows two view landscapes. On the left (Fig.~\ref{fig:view-landscapes}a), a basic responsive visualization with two views (orange and blue areas) and a conventional device breakpoint at a set width (dashed line) is shown.
For any viewport narrower than this threshold, the mobile-focused view (orange) is shown; for viewports wider than the threshold the desktop-focused view (blue) is shown.
On the right (Fig.~\ref{fig:view-landscapes}b), we illustrate a responsive visualization with constraint-based breakpoints. This example has three views (displayed in three different colors) and the constraint-based breakpoints (dashed lines) between them are dynamically generated based on a view stack, view constraints, and a specific dataset. 
Because constraint-based breakpoints use information about the visualized data, view landscapes will look different for different datasets and visualization designs.

\subsection{View Stack \& View Constraints}
\label{sec:breakpoints:stack}

\begin{figure}
    \centering
    \includegraphics[width=\columnwidth]{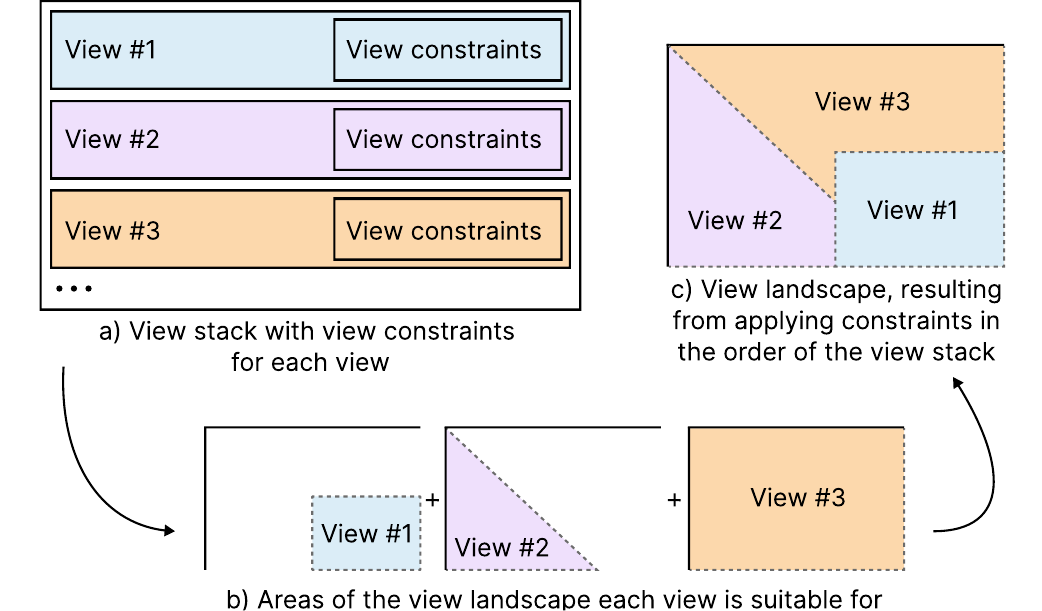}
    \caption{Illustration of how the view stack and view constraints are used to generate the view landscape.}
    \label{fig:viewstack}
\end{figure}

In our approach, we represent a responsive visualization as a stack of multiple \textbf{views}.
Similar to CMV systems~\cite{wangbaldonadoGuidelinesUsingMultiple2000,robertsStateArtCoordinated2007}, each view represents a different visual representation of the data, but where views are \textit{concurrent} in CMV systems, they are \textit{alternatives} in responsive visualization, with only one view shown at a time.
Different views can be entirely different visual representations (e.g., proportional circle map vs. bar chart as in Sec.~\ref{sec:motivating-example}), but they can also be smaller adaptations of a visualization (e.g., proportional circle map vs. Dorling cartogram).
The view stack is \textit{ordered}: the designer's preferred version---typically the most detailed and complete version of the visualization---comes first, followed by other views in descending order of preference. 
Each \textit{view} in the stack has a set of \textbf{view constraints}, which specify what is required for the view to remain legible and usable.

Figure~\ref{fig:viewstack} illustrates this framework using the view landscape. The designer creates the view stack and sets view constraints for each view in the stack (Fig.~\ref{fig:viewstack}a). 
Evaluating the view constraints for a specific dataset yields certain areas of the view landscape that each view is suitable for (Fig.~\ref{fig:viewstack}b). Applying the view constraints in the order given by the view stack is akin to layering these areas over each other and results in the final view landscape, showing the breakpoints between different views (Fig.~\ref{fig:viewstack}c).
For a given dataset and viewport size, our system chooses the most suitable view by (1) applying the view constraints for each view, starting at the top of the stack; (2) if a view violates any of its view constraints, moving on to the next view in the stack; (3) display the first view to fulfill all of its view constraints.

View constraints specify what needs to be given for a visualization to be suitable at a certain size. 
Every visualization design is different and will require different constraints. 
This results in a potentially infinite number of possible view constraints, as well as many different ways of quantifying and implementing these constraints. Therefore, our implemented constraints should not be seen as reference implementations for the different constraint types, but rather examples of possible implementations. 
As demonstrated in Sec.~\ref{sec:examples}, existing constraints can be reused across visualizations, but our design framework also facilitates creating new constraint implementations, which may be required for certain visualization designs, large and complex datasets (requiring highly optimized implementations with low computational complexity), or to achieve a specific visual result.
Through our example visualizations, we identified three general types of constraints:

\textbf{1) Size Constraints}---Scaling a visualization to fit a smaller size, adding new data points with values smaller or larger than what was anticipated in the visualization design, or adding new data points in general can all lead to visual elements becoming too small---such as circles on a proportional circle map, grid cells in a matrix, or spatial units in a choropleth map.
\textit{Too small} can refer to (i) too small to be detectable~\cite{clevelandModelStudyingDisplay1993}, (2) too small to be legible (for textual labels), or (3) too small to be reliably clicked or tapped on for elements with interactive features. This effectively removes the data points encoded by them from the visualization.
For our example visualizations, we implemented five such constraints: \texttt{minAdjacencyMatrixLabelSize} to set a minimum size for the labels (and thereby grid cells) of an adjacency matrix, \texttt{minArcDiagramLabelSize} to set a minimum size for the labels (and thereby distance between nodes) in an arc diagram, \texttt{minAreaSize} to set a minimum size for the smallest spatial unit in a choropleth map,
\texttt{minCircleRadius} to set a minimum size for the circle on a proportional circle map or Dorling cartogram, and
\texttt{minHexSize} to set a minimum size for the hexagons in a hexagonal grid map.

\textbf{2) Overlap Constraints}---When scaling down a visualization using semantic zoom~\cite{perlinPadAlternativeApproach1993} where elements maintain their size but their distance is reduced, or when additional data points are added, elements may start overlapping. Overlap will be exacerbated the more the visualization is scaled down or the more data points are added. For example, in a scatterplot, overplotting will increase the smaller the chart is, in a node-link diagram, dense areas will become increasingly cluttered with overlapping nodes, and in a proportional circle map, circles will increasingly overlap. 
Some amount of overlap is often both acceptable and unavoidable, and its effects can be mitigated through design choices such as applying some transparency to visual elements or scaling elements down. Yet, too much overlap will render the visualization unreadable and make patterns in the data increasingly difficult to identify. 
We implemented one overlap constraint: \texttt{maxOverplotting} to limit the amount of overlap in a scatterplot.

\textbf{3) Aspect Ratio Constraints}---Some visualizations require specific aspect ratios. For instance, an adjacency matrix is always square. Geographic maps also tend to have fixed aspect ratios: a conventional map of the conterminous US is wider than it is tall (landscape format), whereas a conventional map of the UK is taller than it is wide (portrait format), and a map of France is approximately square.
Scaling any of these visualizations to fit a container with a mismatched aspect ratio leads to lots of empty space, indicating inefficient use of space.
Similarly, even charts that have more flexible aspect ratios, like a bar chart, 
may make more efficient use of space if their axes are flipped.
We implemented three aspect ratio constraints: \texttt{maxAspectRatioDiff} to set a maximum difference between the aspect ratio of the container and the visualization (e.g., a geographic map or adjacency matrix with a fixed aspect ratio), and \texttt{minAspectRatio} and \texttt{maxAspectRatio} to specify an overall minimum or maximum aspect ratio for the container; we use this to switch between horizontally and vertically oriented bar charts.

\subsection{Design Workflows With Constraint-Based Breakpoints}
\label{sec:breakpoints:workflows}

\begin{figure}[t]
    \centering
    \includegraphics[width=\linewidth]{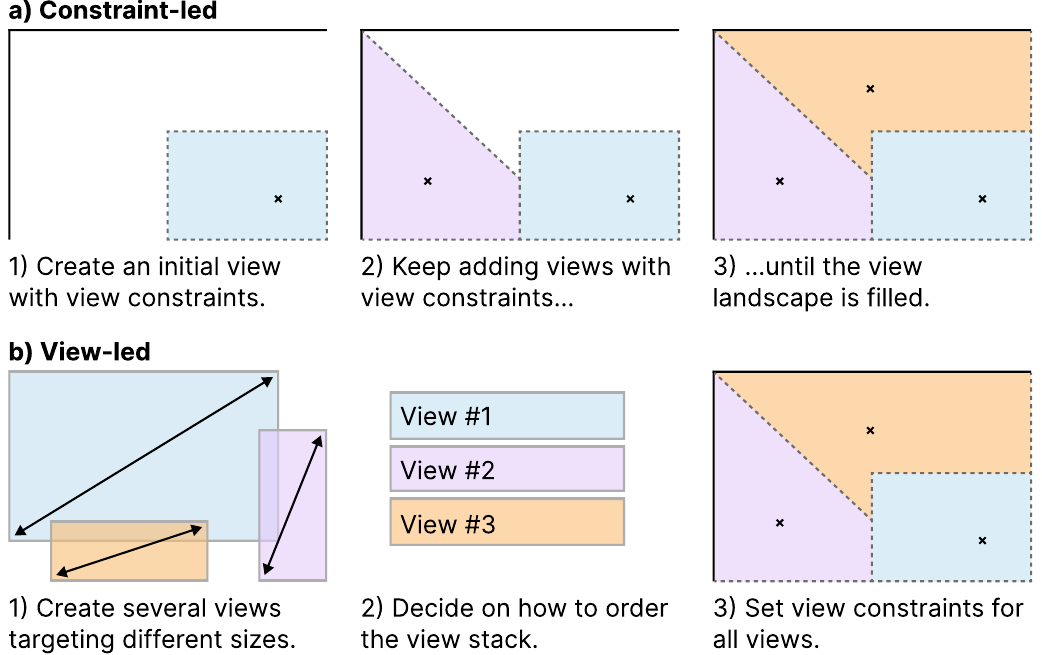}
    \caption{Two workflows for designing with constraint-based breakpoints: In the \conditionsfirst\ process, the designer iteratively adds views with constraints until the view landscape is filled. In the \viewsfirst\ process, all views are designed upfront, then ordered and assigned constraints.}
    \label{fig:workflow}
\end{figure}

We now describe two complementary workflows for designers to extend their existing process to include responsiveness via constraint-based breakpoints. In both workflows, a designer starts with an initial visualization design (view \#1), which will frequently be a large-screen or desktop-focused visualization design, and then proceeds to build the view stack by adding more views and their view constraints. The final result of both approaches is the same, but they cater to different scenarios designers may find themselves in at the start of the process. As a third option, we also discuss mixing these two approaches.

\conditionsfirst\ (Fig.~\ref{fig:workflow}a)---This workflow is best suited for scenarios where designers initially only have a single (ideal) visualization design in mind, and are looking to iteratively discover additional designs suitable for different sizes and aspect ratios. First, the designer examines their initial view \#1 and establishes view constraints that specify when this view becomes unsuitable (Fig.~\ref{fig:workflow}a--1). Second, they design an additional view \#2 intended to solve the readability issue caused by scaling down view \#1 and establish its view constraints (Fig.~\ref{fig:workflow}a--2). Third, they repeat this process and continue adding views with view constraints until the view stack sufficiently fills the view landscape (Fig.~\ref{fig:workflow}a--3). 
The view landscape can act as a design and debugging tool in this process by indicating what area each view is suitable for, and highlighting empty areas in the view landscape that are missing suitable visualization designs.

\viewsfirst\ (Fig.~\ref{fig:workflow}b)---If designers already have several visualization designs for different sizes and aspect ratios in mind, this workflow may be better suited. First, alongside the initial view \#1, the designer creates a set of additional views adapting the visualization for different viewport sizes and aspect ratios (Fig.~\ref{fig:workflow}b--1). Second, the designer orders the views in the view stack. They make this decision based on the specific requirements of their responsive visualization, e.g., 
if minimizing information loss is the most important factor, they might order views according to this metric (Fig.~\ref{fig:workflow}b--2). 
Third, they establish view constraints for each view. The view landscape is then generated by applying the view constraints in the order given by the view stack (Fig.~\ref{fig:workflow}b--3).

\mixed---Often, a mix of these two strategies will be the best workflow. For example, a designer might already have two views in mind with an approximate idea of the parts of the view landscape they fill (\viewsfirst). After creating these two and establishing their view constraints, they can then examine the view landscape to identify areas not covered yet, and fill these iteratively (\conditionsfirst). When creating our example visualizations (Sec.~\ref{sec:examples}), we found that even when we started work on an example with a clear plan for what views to include
(\viewsfirst), we were often surprised by the view landscape, which prompted us to redesign views, add new views, or remove candidate views. For example, Sec.~\ref{sec:examples:popmap} describes how we replaced a view in the \textit{Population Map} example after determining it did not work well for many areas of the view landscape it needed to cover.

\subsection{Implementation}
\label{sec:breakpoints:implementation}

We implemented our design framework as a \textbf{flexible, high-level, light-weight} library. This was driven by a desire to enable \textit{i)} the use of constraint-based breakpoints in tandem with other approaches to responsiveness (Sec.~\ref{sec:rw:designs}) and \textit{ii)} exploration and experimentation with visualization techniques from various areas.
As opposed to declarative grammars~\cite{mcnuttNoGrammarRule2023} such as Vega-Lite~\cite{satyanarayanVegaLiteGrammarInteractive2017} or Cicero~\cite{kimCiceroDeclarativeGrammar2022} for responsive visualization, our implementation does not provide pre-defined specifications of view constraints or breakpoints. 
As constraint-based breakpoints are a novel technique, our implementation instead focuses on enabling designers to explore different designs by facilitating the creation of new views and view constraints. At the same time, views and view constraints are highly reusable and can be recombined into different visualizations once implemented.

We chose to implement our design framework as a \textit{Svelte component}. Svelte~\cite{Svelte} is a JavaScript framework increasingly popular with visualization developers (e.g. \cite{rothschildMakingVisualizationsLiterally2021,stahlSvelteDataViz2022}), often in combination with D3. A core concept in Svelte are \textit{components}: pieces of code that render a specific part of the page. They can be nested and are intended to be reused. We provide a Svelte component that, provided a stack of views and view constraints, evaluates these and renders a responsive visualization.
To create a responsive visualization, designers need to create \textit{components} for each view, as well as a specification that defines their view constraints and the order in which they are stacked.
The component then automatically re-evaluates the view constraints whenever the device context or data changes and updates the view and view landscape as needed. In our interactive demo, we display our example visualizations in containers that can be resized by dragging the bottom-right corner to demonstrate these live updates. In most real-world applications, the container size would instead be determined programmatically, for example by the surrounding layout, or in the case of a full-screen visualization, by the size of the browser window. 
In addition to evaluating view constraints and displaying views, the component also generates an image of the view landscape. All view landscapes shown in Sec.~\ref{sec:examples} were generated this way, and they are also shown as an optional overlay in our online demo. 

For our example visualizations, we implemented six visualization components, which enabled us to create a total of 13 different \textit{views} across four example responsive visualizations: 

\begin{itemize}
    \item A \textbf{Vega-Lite wrapper}: passing any valid Vega-Lite~\cite{satyanarayanVegaLiteGrammarInteractive2017} specification to this component will render it as a view. The component implements view constraints on overplotting in scatterplots, cell size in heatmaps, and aspect ratio for vertical vs. horizontal bar charts.
    \item A \textbf{NetPanorama wrapper}: NetPanorama~\cite{scott-brownNetPanoramaDeclarativeGrammar2023} is a declarative grammar for network visualization. Passing any valid NetPanorama specification to this component will render it as a view. The component implements view constraints on the label size in arc diagrams and adjacency matrices. 
    \item Four \textbf{custom D3-based visualizations}: we used D3.js~\cite{bostockDataDrivenDocuments2011} to create four visualization components, a choropleth map, a hexagon map (using the \texttt{d3-hexjson} module~\cite{hawkinsD3hexjson2022}), a proportional circle map/cartogram, and a waffle chart, each with one or more view constraints. 
\end{itemize}

The following section provides additional detail on the implemented views and constraints. 
The full code for our implementation and example visualizations and  detailed documentation of all implemented views and constraints is available via our companion website: \sitelink.
\section{Example Visualizations}
\label{sec:examples}

\begin{table}
    \centering
    \caption{Overview of example visualizations}
    \setlength\tabcolsep{0.1cm}
\setlist[itemize]{nosep, leftmargin=*, labelindent=0cm,labelsep=0.1cm}
\scriptsize
\begin{tabular}{>{\raggedright\arraybackslash}m{1.1cm} >{\raggedright\arraybackslash}m{1.2cm} >{\raggedright\arraybackslash}m{2.3cm} >{\raggedright\arraybackslash}m{2.3cm} >{\raggedright\arraybackslash}m{1cm}}
\textbf{Example} & \textbf{Data types} & \textbf{Visualization types} & \textbf{Constraint types} & \textbf{Workflow} \\\toprule
\textbf{Population map} & 
Geographic, numerical & 
\begin{itemize}
\item Proportional circle map
\item Dorling cartogram
\item Horizontal bar chart
\item Vertical bar chart
\vspace*{-\baselineskip}
\end{itemize} &
\begin{itemize}
\item Size of circles
\item Number of bars
\item Length of bars
\item Aspect ratio of screen
\vspace*{-\baselineskip}
\end{itemize} &
Constraint-led
\\\midrule
\textbf{UK Election map} &
Geographic, categorical &
\begin{itemize}
\item Choropleth map
\item Hexagonal grid map
\item Waffle chart
\vspace*{-\baselineskip}
\end{itemize} &
\begin{itemize}
\item Size of spatial units
\item Size of hexagons
\item Size of squares
\item Aspect ratio of screen
\vspace*{-\baselineskip}
\end{itemize} &
Constraint-led
\\\midrule
\textbf{Network} &
Network data (undirected) &
\begin{itemize}
\item Arc diagram
\item Node-link diagram
\item Adjacency matrix
\vspace*{-\baselineskip}
\end{itemize} &
\begin{itemize}
\item Size of labels
\vspace*{-\baselineskip}
\end{itemize} &
View-led
\\\midrule
\textbf{Scatterplot} & 
Numerical, tabular &
\begin{itemize}
\item Scatterplot
\item Heatmap
\vspace*{-\baselineskip}
\end{itemize} &
\begin{itemize}
\item Overlap of points
\vspace*{-\baselineskip}
\end{itemize} &
View-led\\
\end{tabular}

    \label{tab:examples}
\end{table}

In this section, we describe four example visualizations with a total of nine variations using different datasets or data subsets and different constraints. 
We describe our design process and rationale for a first example in detail to 
illustrate how the workflows described in Sec.~\ref{sec:breakpoints:workflows} unfold and how our toolkit supports iterating on the choice of views as well as the constraint-based breakpoints separating them.
Additionally, we provide more concise descriptions of a further three examples to illustrate a wider variety of data types, visualization designs, constraint types, and design workflows;  
Table~\ref{tab:examples} provides an overview of how the four examples cover this variety.
All visualizations and their variations are available in our interactive online demo: \sitelink.

\subsection{Detailed Design Process: Visualizing Global Population}
\label{sec:examples:popmap}

This visualization compares the population of countries and territories around the world 
using a dataset of estimated population counts for 258 countries/territories in 2017~\cite{naturalearthAdminCountriesFree}.

\begin{figure}
    \centering
    \includegraphics[width=\linewidth]{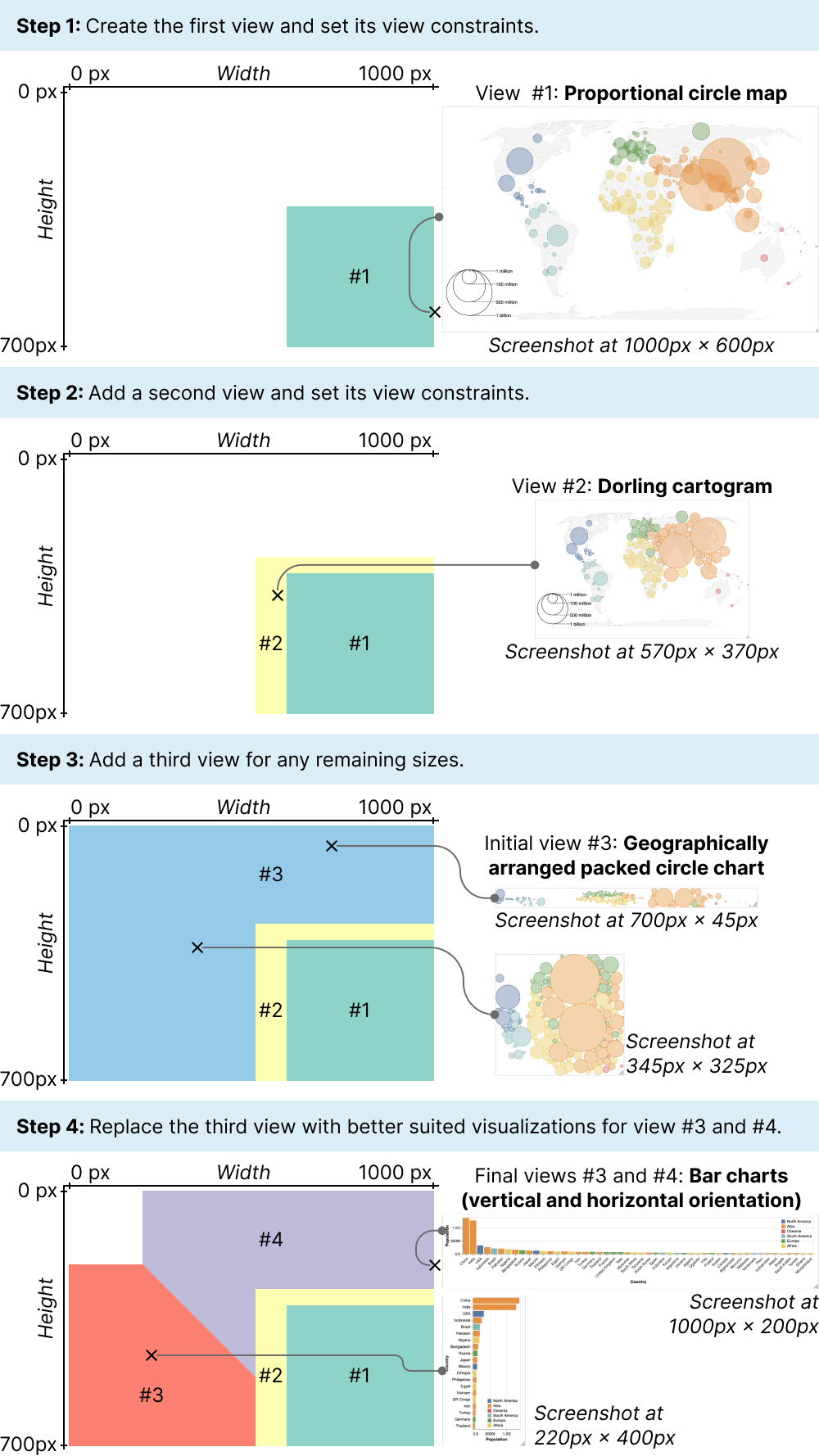}
    \caption{Step-by-step overview of the design process for the \textit{Population Map} example: we generally followed a \conditionsfirst\ approach, but decided to modify view \#3 after examining the final visualization. The interactively resizable demo is available online: \demolink{population-map}.}
    \label{fig:pop-map-steps}
\end{figure}

\textbf{\conditionsfirst, Step 1: Create view \#1 with view constraints---}We designed this example with a \conditionsfirst\ approach. Map-based visualizations are the obvious choice for visualizing geographic data and making patterns and distributions visible, so we started with a proportional circle map (Fig.~\ref{fig:pop-map-steps}--Step 1). This visualization clearly shows all population counts in their geographic context by visualizing each population count as a proportionally scaled circle located at the center of each country or territory. 
The main issue when scaling down a proportional circle map (assuming the entire visualization is scaled geometrically) is that many of the circles become too small to be detectable. We considered it acceptable for \textit{some} of the smallest populations to not be detectable; due to the large range of population counts in the dataset (ranging from 140 to 1.4 billion), this is essentially unavoidable. Assigning just a single pixel to the smallest circle representing a population of 140 would require the largest circle to have a diameter of over 3,500px. Many of the polygons mapping out smaller countries on the underlying map also become so small that they are no longer detectable, but we considered this acceptable as each country's location would still be marked by a circle. We therefore set an arbitrary, but informed, view constraint that no more than 10\% of all circles may be smaller than 2px in diameter.

\textbf{\conditionsfirst, Step 2: Add view \#2 with view constraints---}For cases where the proportional circle map is too large, we added a Dorling cartogram~\cite{dorlingAreaCartogramsTheir1996} as a second view (Fig.~\ref{fig:pop-map-steps}--Step 2). This is similar to a proportional circle map but positions circles such that they do not overlap, while keeping them as close as possible to the center of the country on the underlying map. It adds some distortion to the geographic locations of most circles, but in turn allows us to use slightly larger circles than in a proportional circle map since overlap is no longer an issue.
Scaling down a Dorling cartogram is similar to scaling down a proportional circle map, and the same considerations in terms of what we consider an acceptable loss of information apply. Therefore, we set the same view constraint for the Dorling cartogram: no more than 10\% of all circles may be smaller than 2px in diameter.

\textbf{\conditionsfirst, Step 3 (First attempt): Add view \#3---}For any remaining smaller sizes, we initially added a custom, geographically arranged packed circle chart (Fig.~\ref{fig:pop-map-steps}--Step 3). 
Our goal with this was to preserve the visual encoding of population counts to circle size, and maintain \textit{some} geographic context.
This custom chart tries to maintain the general geographic layout of the circles, but nonetheless heavily compromises geographic relationships. 
After implementing this, we realized that the packed circle chart did not work well for many of the sizes and aspect ratios it would need to cover (Fig.~\ref{fig:pop-map-steps}--Step 3, blue areas in the view landscape). Many circles were too small since the size of the largest circles was limited by needing to fit into a narrow rectangle, even if there was spare space overall. Additionally, the geographic patterns, which we had intended to somewhat preserve with this solution, were not clearly decipherable at most sizes and aspect ratios. We therefore decided to shift to a different visualization design.

\textbf{\conditionsfirst, Step 3 (Second attempt): Review and replace third view---}In our second attempt at designing a third view, 
we opted for a horizontal and a vertical bar chart (implemented with Vega-Lite).
We set the vertically oriented bar chart as view \#3 with a view constraint requiring the aspect ratio of the container to be larger than 1, i.e., wider than it is tall, and the horizontally oriented bar chart as view \#4 shown in all other cases---these could be swapped and the constraint reversed for an identical outcome. The bar charts are sorted by population and filtered to show only as many bars as can fit with all labels remaining legible and not overlapping. Alternatively, the bar chart could be made scrollable to show more data. Across all views, the circles/bars are colored by continent, which provides some basic geographic context in the bar chart. 

\begin{figure}
    \centering
    \includegraphics[width=\linewidth]{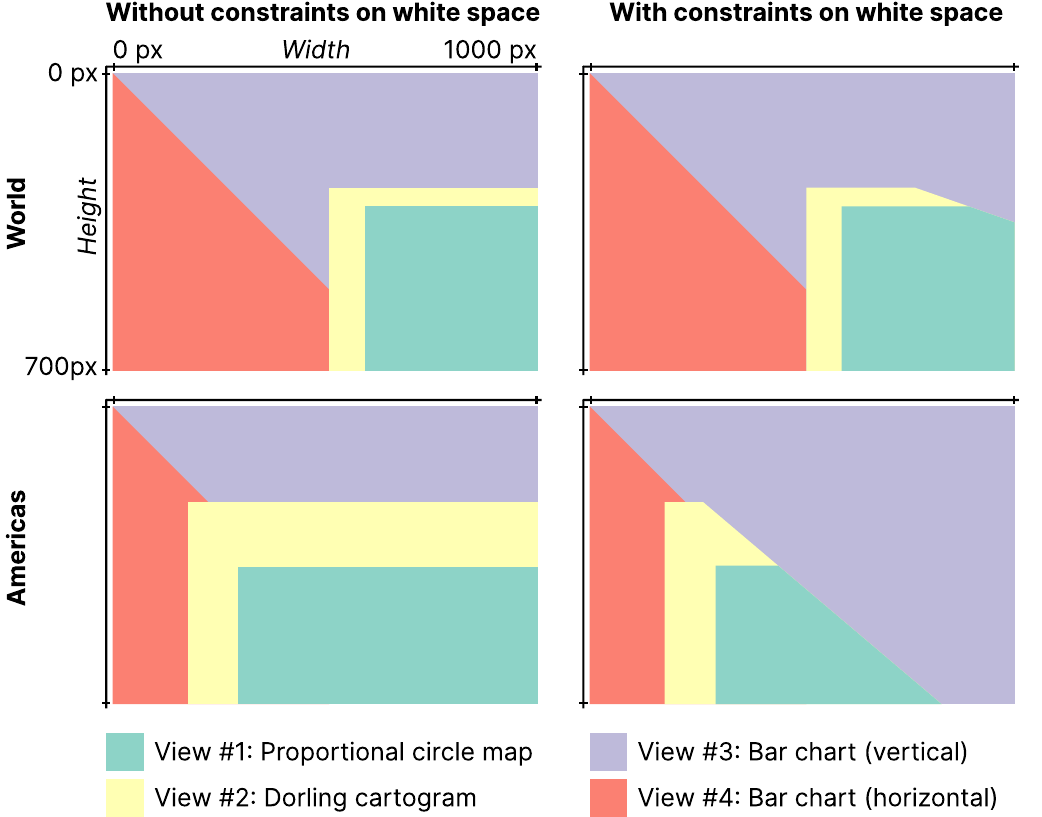}
    \caption{Comparison of view landscapes when varying data and constraints: Views \#1 (green) and \#2 (yellow) are suitable for much smaller screens if only the Americas (bottom row) rather than the whole world (top row) are displayed. Introducing constraints on the amount of empty space that is permitted around the map only has a small effect on the world map (top right), but significantly changes the view landscape for the Americas (bottom right).}
    \label{fig:comparison_vl_population}
\end{figure}

\textbf{Considering additional view constraints---}A secondary issue is that when maps are scaled to fit a container with a much wider or taller aspect ratio, there will be significant white space around the map. This can simply be aesthetically undesirable, but also indicates inefficient use of space.
Figure~\ref{fig:comparison_vl_population} illustrates how adding a constraint requiring the aspect ratio of the container to be no more than 50\% wider or taller than the map would affect the view landscape.
We can see that the new constraint cuts into the area covered by views \#1 and \#2, particularly for the data subset described below.
However, in deciding whether to actually add this constraint, we need to consider the alternative: the proportional circle map and Dorling cartogram have similar aspect ratios, so showing one over the other does not reduce empty space. The bar chart removes nearly all geographic context and provides less information than views \#1 and \#2, so choosing it over accepting empty space around the map is a poor trade-off.

\textbf{Varying the dataset---}Finally, we created an additional version showing a subset of the original global dataset: North and South America. 
This data subset has a smaller range of values and a narrower map aspect ratio.
Since constraint-based breakpoints seamlessly adapt to these changes, we only have to make some minor tweaks to the visualization: adapting the parameters of the map projection and slightly increasing the scale of the circles in views \#1 and \#2 to make better use of the space with a smaller dataset.
Fig.~\ref{fig:comparison_vl_population} illustrates the effect of switching to this data subset: views \#1 and \#2 are shown for much smaller widths and the additional aspect ratio constraint has a much more extreme effect.

\subsection{Further Examples}
\label{sec:examples:additional}

This section describes three additional examples that demonstrate designing with constraint-based breakpoints for categorical geographic data, multivariate data, and networks. 

\begin{figure}
    \centering
    \includegraphics[width=\linewidth]{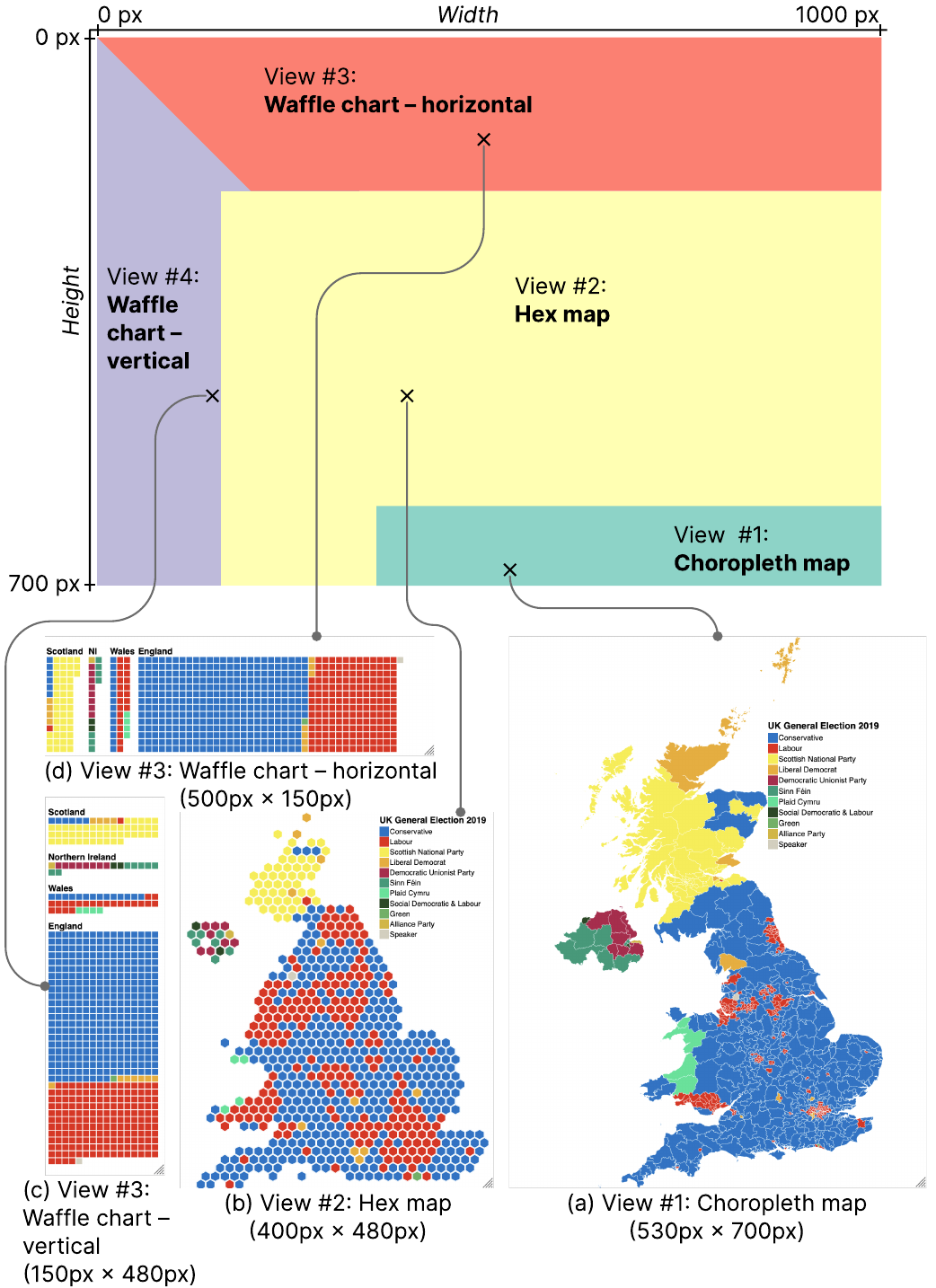}
    \caption{Overview of the \textit{UK Election} example: The view landscape shows which view is shown at each screen size. Each view (\#1 choropleth map, \#2 hex map, \#3 and \#4 waffle charts) is scalable; we show screenshots taken at specific sizes. The interactively resizable demo is available online: \demolink{uk-election}}
    \label{fig:uk_small}
\end{figure}

\textbf{Categorical Geographic Data: Choropleth Maps, Hex Maps, \& Waffle Charts---}This example visualizes the results of the 2019 General Election in the UK using three different views. 
The dataset~\cite{bakerGeneralElection20192020} contains categorical data about which political party was elected in each of the UK's 650 constituencies, with 10 different parties winning in at least one constituency.
This visualization could be used in an online news context, where readers use a large variety of devices to access the page. The design of the maps in this example was inspired by the BBC's reporting on this election~\cite{bbcResults2019General2019}.
Our first view is a choropleth map using the Albers equal-area conic projection~\cite{snyderMapProjectionsWorking1987} (Fig.~\ref{fig:uk_small}a), showing each constituency in its geographic context while preserving the relative size of areas. 
However, due to constituencies being mainly designed to have equal population~\cite{alvanidesZoneDesignPlanning1999}, their geographic areas vary considerably, leading to small urban constituencies being undetectable when scaled down.
To account for this, we set a view constraint requiring a minimum size of 2px for the smallest constituency.
We then add a second view (Fig.~\ref{fig:uk_small}b): a hexagonal grid map (`hex map') representing each constituency as a hexagon of the same size. 
This presents a good compromise between ensuring each constituency is visible and preserving national and local geographic relationships to a degree.
Since all hexagons are the same size, we set a view constraint requiring a minimum width of 5px for each hexagon, equivalent to an area of about 22px.
This mainly leaves small sizes and aspect ratios unsuitable for a map of the UK to consider.
For views \#3 and \#4, we therefore opt for a waffle-chart-like visualization in a vertical and a horizontal version (Fig.~\ref{fig:uk_small}c+d). This design is very flexible in terms of rearranging the squares representing each constituency, meaning it can be scaled to small sizes and extreme aspect ratios while all constituencies remain visible. However, despite grouping constituencies by nation (Scotland, England, Northern Ireland, Wales), it sacrifices nearly all geographic information.
On the horizontal version of the waffle chart (view \#3), we set a constraint requiring the aspect ratio to be smaller than 1, i.e., landscape format. View \#4, the vertical version of the chart, is shown in all other cases. For both of these views, we additionally require each square to be at least 2px wide (equivalent to an area of 4px), so that they are visible, if very small.

\begin{figure}
    \centering
    \includegraphics[width=\linewidth]{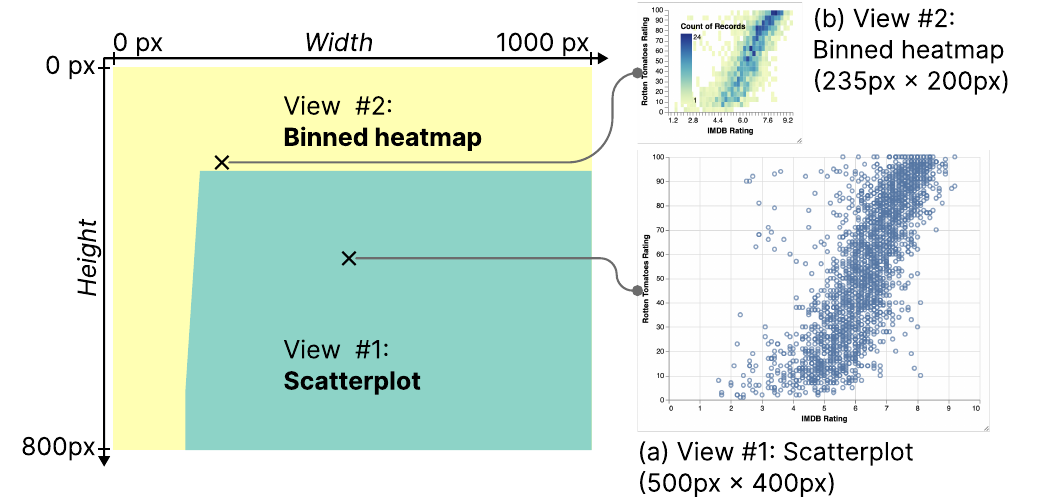}
    \caption{Overview of the \textit{Scatterplot} example: The view landscape shows which view is shown at each screen size. Each view (\#1 scatterplot, \#2 binned heatmap) is scalable; we show screenshots taken at specific sizes. The interactively resizable demo is available online: \demolink{scatterplot}}
    \label{fig:scatter_small}
\end{figure}

\textbf{Quantitative Bivariate Data: Scatterplots and Binned Heatmaps---}Scatterplots often have issues with overplotting, potentially leading to false information being perceived~\cite{chenUsingAnimationAlleviate2018}.
Small design changes, such as applying transparency to visual marks and reducing their size, can help alleviate issues with overplotting. In our example visualization (Fig.~\ref{fig:scatter_small}) using a common sample dataset about movie ratings~\cite{vegaVegaDatasets}, we combine this with setting a view constraint on the scatterplot that limits the amount of permitted overlap.
We quantify overplotting by computing overlap between any two marks as a value between 0 and 1, where 0 is no overlap, and 1 indicates identical positions. 
We calculate total overlap as the proportion of this value compared to all marks overlapping. This value is limited to no more than $0.009$ in our example. 
As an alternative, view \#2 shows the same data in a binned heatmap (Fig.~\ref{fig:scatter_small}b). 

\begin{figure}
    \centering
    \includegraphics[width=\linewidth]{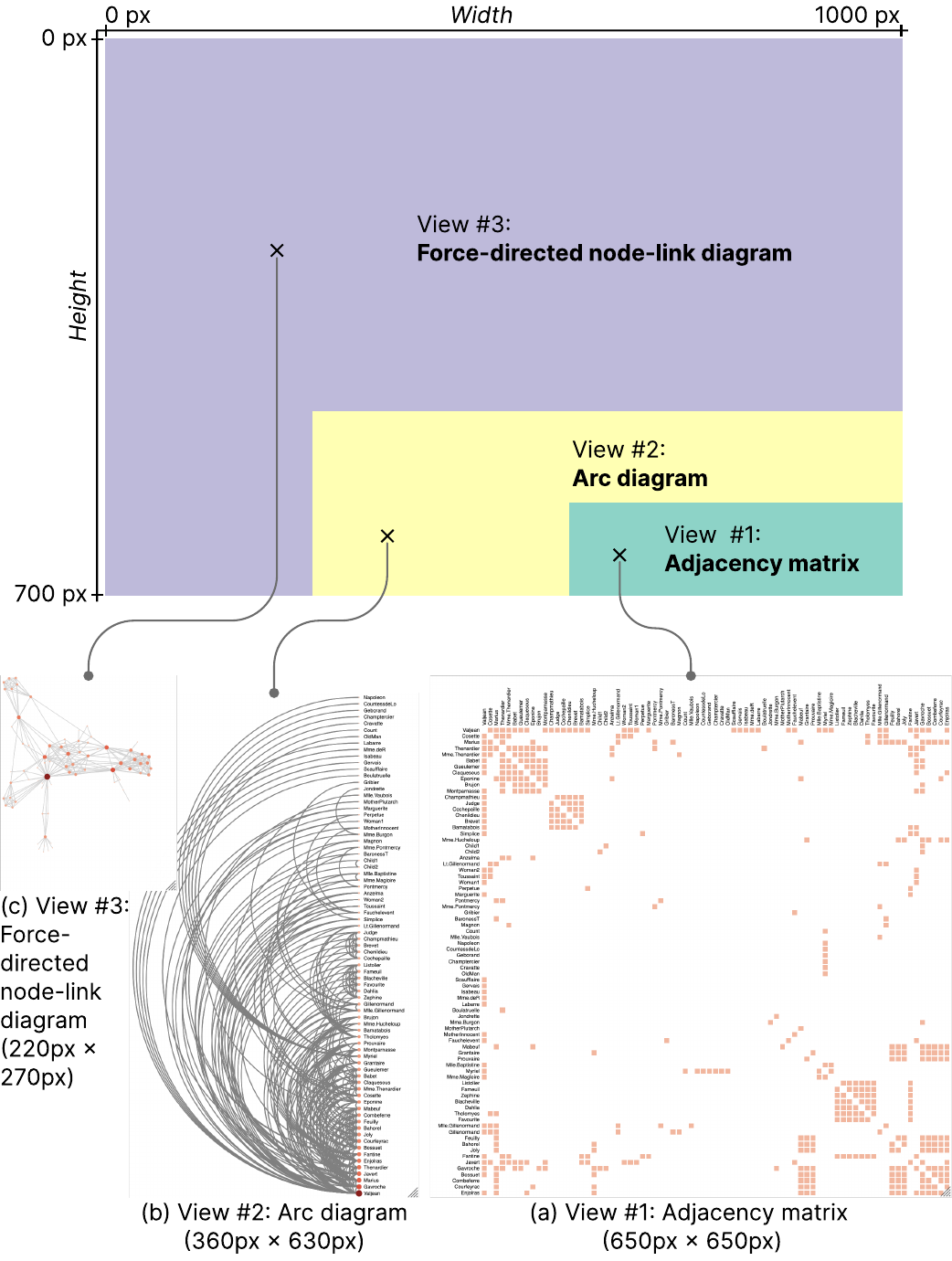}
    \caption{Overview of the \textit{Networks} example: The view landscape shows which view is shown at each screen size. Each view (\#1 adjacency matrix, \#2 arc diagram, \#3 node-link diagram) is scalable; we show screenshots taken at specific sizes. The interactively resizable demo is available online: \demolink{networks}}
    \label{fig:network_small}
\end{figure}

\textbf{Graphs and Networks: Node-Link Diagrams, Arc Diagrams, \& Adjacency Matrices---}While less popular in online news media, network visualizations are crucial to many multi-view analytic applications. 
For dense networks, networks with link attributes, or tasks related to clusters, adjacency matrices (Fig.~\ref{fig:network_small}a) are a good option, so we choose this visualization as view \#1. For an adjacency matrix to be readable, we mainly need to ensure that the row and column labels are legible. Hence, we set a constraint requiring the height (for row labels) and width (for column labels) to be 6px or more. For the Les Misérables character co-occurrence dataset~\cite{vegaVegaDatasets}, which has 77 nodes, shown in Fig.~\ref{fig:network_small}, this results in the matrix only being shown at sizes above 600px in width and height. 
Now, if we need some of our screen space for other visual content (additional visualizations or other UI elements) or use a smaller device, we may need to rescale our visualization but we do not want to lose information about specific nodes and links through aggregation~\cite{elmqvistZAMEInteractiveLargeScale2008}.
Therefore we introduce an arc diagram as an alternative visual representation for view \#2 (Fig.~\ref{fig:network_small}b). Compared to the adjacency matrix, this shows connections less efficiently, but avoids aggregation and maintains the network's topology. Despite applying the same view constraint requiring the labels to be 6px or more in height, the view landscape shows that this view is suitable for slightly smaller heights, and much smaller widths.
Last, to account for the remaining areas of the view landscape, we introduce a force-directed node-link diagram (view \#3). This can make efficient use of the available screen space by distributing nodes inside the entire available space. 
\section{Discussion}
\label{sec:discussion}

Across a range of use cases and data types, we demonstrated how responsive visualizations can utilize constraint-based breakpoints and how designers can take advantage of these.
In contrast to the dominant top-down approach of developing visualizations to fit pre-defined  breakpoints based on standardized width thresholds, constraint-based breakpoints provide fine-grained control across a wider range of devices, and  open up new design possibilities by accounting for changes to values and variation in the underlying data.
We show how constraints based on visual element size, overlap and aspect ratio for particular visualization designs can be used in responsive visualization.
Our toolkit and examples reveal the scope of these designs and allow us to explore the effects of different constraints as we fill the view landscape.

\subsection{Limitations and Extensions}

Our examples in Sec.~\ref{sec:examples} also show limitations in our approach that inspire future research.

\textbf{Strategies for adapting views---}We have focused on different visualization \textit{designs} as solutions for different sizes. However, several of our examples in Sec.~\ref{sec:examples} could have equally been addressed by other view-changing strategies. Our examples already make use of some common strategies such as aggregation (our scatterplot is replaced by a binned heatmap) and filtering (the bar chart in our population map only shows the most populous countries). Aggregation combines data elements into groups and visualizes information about these groups, which can be efficiently automated for some data types~\cite{elmqvistZAMEInteractiveLargeScale2008}. Other options are suggested by Horak~et~al.~\cite{horakResponsiveVisualizationDesign2021} and by Kim~et~al.'s responsive design patterns~\cite{kimDesignPatternsTrade2021}, such as zooming in on a map based on a viewer's geographic position. 
For exploratory or analytics-focused visualizations where more user input is expected, there might be approaches to learn from a user's journey~\cite{sperrleLotsePracticalFramework2023} to help suggest views that minimize disruption. 

\textbf{Creating a consistent set of views---}These different approaches to changing views raise important questions around how to design a \textit{meaningful ensemble of designs} for different views---beyond adapting an individual visualization design, what do we need to consider when combining different visualization designs into one coherent responsive visualization?
In Sec.~\ref{sec:examples}, we explain our design rationale for each example visualization, considering factors such as maintaining consistent visual encodings across views, preserving as much information in the visualization as possible, or choosing visualization designs that maintain geographic patterns. User studies could explore how different combinations of views in responsive visualizations are perceived and reveal what factors, such as consistency, information density, or familiarity of a visualization design, matter most in designing effective responsive visualizations.

\textbf{Smoothing view transitions---}In many usage scenarios for responsive visualization, users will only ever see one view that fits the device they are using. However, users may see multiple views if they choose to rotate their device, resize the browser window, or continue reading on a different device. And scenarios such as interactively resizable dashboards and systems combining multiple responsive visualizations could lead to users seeing potentially many different views. In these scenarios, it becomes more important to present smooth responses that support rather than disrupt tasks when encountering a breakpoint. 
We already discussed choosing consistent visual encodings, which may be perceived as less abrupt and help users keep track of visual elements.
Alternatively, or in addition, we could consider introducing animated transitions between views. There are many validated designs and implementations for animated transitions between visualizations that responsive design could draw on (e.g.,~\cite{heerAnimatedTransitionsStatistical2007,kimGeminiGrammarRecommender2021,chalbiCommonFateAnimated2020,chevalierNotsoStaggeringEffectStaggered2014,duTrajectoryBundlingAnimated2015}), including many techniques targeting more specific use cases such as transitioning from a map to a cartogram~\cite{slingsbyConfiguringHierarchicalLayouts2009}, or transitioning between different types of unit visualizations~\cite{parkAtomGrammarUnit2018,druckerUnifyingFrameworkAnimated2015}.

\textbf{Default constraints---}Our approach currently relies on manually choosing and quantifying constraints for each visualization design, as we demonstrate in Sec.~\ref{sec:examples}. 
Empirical user-centered studies on defining constraints could validate the approach further and yield well-justified default constraints for different visualization types.
Formalized constraints have previously been explored in visualization in the context of automated visualization design~\cite{moritzFormalizingVisualizationDesign2019}, offering a collection of constraints that could act as a starting point for a responsiveness-focused collection.
Such a library of default constraints, possibly in the form of a declarative grammar that complements or extends an existing visualization grammar such as Vega-Lite~\cite{satyanarayanVegaLiteGrammarInteractive2017}, could enable designers to quickly configure responsive visualizations with very little code, 
or even serve as the basis for a UI-based authoring tool.
This would significantly lower the time investment and coding skills required to publish a fully responsive visualization,
thus streamlining the design process and making responsive visualization accessible to less experienced visualization creators.

\textbf{Additional aspects of responsiveness---}We demonstrated how constraint-based breakpoints simultaneously account for changes to the screen size and the visualized data. However, Horak~et~al.~\cite{horakResponsiveVisualizationDesign2021} describe many more factors to consider for responsiveness. More advanced device factors, such as non-rectangular screens (e.g., round smartwatches) or different interaction modalities could easily be accounted for with constraint-based breakpoints. However, constraints depend on being \textit{measurable} on the user's device, making human factors such as the user's visualization literacy, environmental factors such as crowdedness, and usage factors such as a user's posture difficult to account for---at least without major privacy concerns. Another caveat is that the two-dimensional view landscape visualization we introduce does not easily scale up to display additional factors.

\textbf{Giving users agency over changing views---}The different views in our responsive visualization examples support different tasks to varying degrees of success. In scenarios where users interactively adapt the size of the visualization or modify the data, a designer may want to give users more agency over switching views.
Since constraint-based breakpoints are linked to specific issues in the visualization, they open up the opportunity to give viewers precise explanations of why another view might be more suitable, letting them make the choice. In this scenario, encountering a breakpoint might trigger a popup or banner that informs the viewer that, e.g., `This scatterplot is very overplotted and you may not be able to see patterns well. Switch to binned heatmap?', thereby enabling the user to make an informed decision. This approach has potential far beyond responsive visualization, e.g., to prevent misinterpretations and deception~\cite{chenUsingAnimationAlleviate2018}.

\subsection{How Can Constraint-Based Breakpoints Change the Way We Think About Responsive Visualization?} 

Many technological and conceptual innovations disrupt the work and routine processes of designers and other people. What can lead to opportunities in the long run can cause friction in a short term. For example, while introducing interactive data comics to comic artists~\cite{wangInteractiveDataComics2022}, artists commented on the huge space of possibilities they had never thought about before and that they needed to explore and get familiar with the novel concepts to lead them to impact. 
Constraint-based breakpoints will almost certainly require a similar investment in exploration by practitioners and designers to understand the implications for responsive design. This includes exploring questions such as:
\textit{Which visualizations cause the least disruption across different views if a change in visualization is necessary}? 
\textit{What are meaningful constraints to (automatically) assess the legibility and usability of a visualization?}
\textit{What messages and information do I want to preserve across views?}
\textit{How many views can and should a view landscape contain, and how do viewers react to and navigate through these responsive view landscapes?}
\textit{What do I know about my audience, their interests in the data and their visual(ization) literacy, and how can I account for their diverse interests and needs in my responsive design process?}
These and similar questions help discuss the future of visualization and analytics across scenarios that involve an increasing number of devices, applications, people, and tasks. 

\section{Conclusion}
\label{sec:conclusion}

This paper introduced constraint-based breakpoints for responsive visualization, a novel concept to facilitate adapting visualizations across device contexts. 
Our work on constraint-based breakpoints demonstrates the potential of reframing the challenges of responsive visualization. Where prior work has mostly focused on how visualizations might be adapted for different devices, we advocate for seeing responsive visualizations as not just a collection of different versions of a visualization, but a system where the mechanics of choosing which of these versions is displayed are essential. 
As is often the case for new ideas, there are many open questions to be considered in future work, such as how to develop meaningful compositions of different views for a variety of use cases, how to prevent users from experiencing abrupt transitions, and how to further support designers' workflows. 
Answering these questions will require time and a combination of different approaches, but our exploration of possibilities suggests this may be fruitful and lead to novel strategies and tools.

% \section*{Supplemental Materials}
% \label{sec:supplemental_materials}
% Our supplemental materials include (1) our companion website providing interactively resizable versions of the example visualizations described in this paper, accessible at \sitelink{}, (2) additional detail about our implementation, (3) a video demonstrating our interactive example visualizations (both also included on the website).

\section*{Acknowledgments}
We would like to thank the anonymous reviewers for their thoughtful and constructive comments, particularly those who reviewed multiple iterations of this paper.
Sarah Schöttler is supported by an EPSRC grant (project ref. EP/T517884/1).

% IEEEtranS is sorted, IEEEtran is ordered by appearance
\bibliographystyle{IEEEtranS}
\bibliography{responsive_geovis}

\section*{Biography Section}
% If you have an EPS/PDF photo (graphicx package needed), extra braces are
%  needed around the contents of the optional argument to biography to prevent
%  the LaTeX parser from getting confused when it sees the complicated
%  $\backslash${\tt{includegraphics}} command within an optional argument. (You can create
%  your own custom macro containing the $\backslash${\tt{includegraphics}} command to make things
%  simpler here.)
 
% \vspace{11pt}

% \bf{If you include a photo:}\vspace{-33pt}
% \begin{IEEEbiography}[{\includegraphics[width=1in,height=1.25in,clip,keepaspectratio]{fig1}}]{Michael Shell}
% Use $\backslash${\tt{begin\{IEEEbiography\}}} and then for the 1st argument use $\backslash${\tt{includegraphics}} to declare and link the author photo.
% Use the author name as the 3rd argument followed by the biography text.
% \end{IEEEbiography}

% \vspace{11pt}

% \bf{If you will not include a photo:}\vspace{-33pt}
% \begin{IEEEbiographynophoto}{John Doe}
% Use $\backslash${\tt{begin\{IEEEbiographynophoto\}}} and the author name as the argument followed by the biography text.
% \end{IEEEbiographynophoto}

\vspace{-33pt}
\begin{IEEEbiographynophoto}{Sarah Schöttler} 
is a PhD candidate at the VisHub lab at the University of Edinburgh. Her research explores responsive visualization with a focus on geographic data. 
%She has an MSc in Design Informatics, also from the University of Edinburgh. 
She has previously published a survey of geographic network visualization techniques, which sparked her interest in applications for unusual geographic visualizations---such as responsive visualization. 
\end{IEEEbiographynophoto}
\vspace{-33pt}
\begin{IEEEbiographynophoto}{Jason Dykes} is a Professor of Visualization and directs the giCentre at City, University or London.
A National Teaching Fellow of the UK HE Academy (2005), he designs interactive visualization techniques that reveal complexity and structure in data. His 20+ papers in IEEE Transactions on Visualization and Computer Graphics describe diverse and innovative techniques for data visualization, in a wide range of domains, and creative methods for developing them.
\end{IEEEbiographynophoto}
\vspace{-33pt}
\begin{IEEEbiographynophoto}{Jo Wood} is a Professor of Visual Analytics at the giCentre, City University of London. He has research interests in novel forms of geovisualization, in narrative design to support visual analytics and communication and in computational thinking in pedagogy. He has over 100 peer reviewed publications in the fields of visualization and education and has been a program and organising committee member for several IEEE venues including VIS and EuroVIS.
\end{IEEEbiographynophoto}
\vspace{-33pt}
\begin{IEEEbiographynophoto}{Uta Hinrichs} is a Reader in Data Visualization at the University of Edinburgh. Her research at the intersection of visualization and HCI is motivated by an urge to understand and facilitate how people engage with information in physical and digital spaces. Her research is highly interdisciplinary, combining methods from design, computer science and the humanities, also visible in her range of publications that span the fields of visualization, HCI, and the (digital) humanities.
\end{IEEEbiographynophoto}
\vspace{-33pt}
\begin{IEEEbiographynophoto}{Benjamin Bach} is a Research Scientist at Inria (France) and Reader in Design Informatics and Visualization at the University of Edinburgh. His research investigates effective and efficient data visualizations, interfaces, and tools for data analysis, communication, and education in visualization. \texttt{\url{benjbach.me}}.
\end{IEEEbiographynophoto}

\vfill

\end{document}